# Passenger life-saving in a badly damaged aircraft scenario

## Alexander Bolonkin
C&R, 1310 Avenue R, #F-6, Brooklyn, NY 11229, USA
T/F 718-339-4563, aBolonkin@juno.com, http://Bolonkin.narod.ru

## Abstract

Offered is a new method for saving passenger lives in any catastrofic situation, including total failure of aircraft control, extreme damage and loss aircraft wings, tail, breakdown all propelling engines, etc. It shown here that previous works which have proposed using only parachutes are useless because their proposers failed to consider the likely overload of the parachute jerk stress (at the moment of parachute release) and the impact of aircraft on Earth surface. These jeck and impact destroy aircraft and kill passengers.

Offered is a connected series of related technical innovations which overcome these obvious difficalties and allow for a soft, near zero speed landing in any topographically suitable place, allowing potential to save aircraft. This method may be applied to all existing airplanes and increases their weight only about $1.5 \div 2.5\%$. Also, the method may be used for vertically landing the already built aircraft, for example, when any runway is damaged or would become overloaded.

------------------

## Introduction

Relative deaths of air passengers are sometimes more than the death rate due to railroads or ships. Especially, the probability of death in an aircraft catastrophe has increased since the start of the War on Terror after 9/11/01. However, people cannot abandon $21^{st}$ Century aviation. Aviation transport is the most comfortable, quick and sometimes only means to reach many distant Earth regions. So, engineers must find the best method of saving passenger lives when aircraft are damaged and even when a full loss of control happens.

Author offers some unique innovations and shows that offered devices installed on any current aircraft can save passenger lives in any airplane damage include wing failure, tail, engine, control and diving from high altitude. The relative weight of offered devices is only $1.5 \div 2.5\%$ of aircraft weight.

Aircraft catastrophes are very different from sea or railway catastrophes [1]. In any ship catastrophe there is time to call land bases or other ships for help. Ships have lifeboats. In any railway catastrophe the most part of passengers (especially in last cars) are safe. In marked contrast, aviation catastrophes usually result in all passengers being killed as the aircraft plummets from the sky. About 40 passengers out of 1 million are fatalities in every flight (It is without terrorist attack. The terrorist attack increases it in 4 times). American Airlines Flight 587 crashed in Queens, NY, killing 256 people. Aircraft crashes dissuade people from flying as passengers.  On September 11, 2001 shows the big, fully fueled commercial aircraft are potential super-bombs that cans kill not only all 200 - 300 passengers but thousands of people on the ground too.

**1. Aircraft accidents**. Approximately 80 percent of all aviation accidents occur shortly before, after, or during takeoff or landing, and are typically the result of human error and/or ignored technical problems within an aircraft; mid-flight disasters are rare but not entirely uncommon. Among other things, the latter have been caused by smuggled bombs as in the 1988 Lockerbie incident, mid-air collisions such as in the 2002 Überlingen crash or in cases of (purportedly) mistaken identity where civilian aircraft were shot down by military (compare Korean Air Flight 007).

An accident survey of **2,147** aircraft accidents from 1950 through 2004 determined the causes to be as follows:

- 45%: Pilot error
- 33%: Undetermined or missing in the record
- 13%: Mechanical failure
- 7%: Weather
- 5%: Sabotage (bombs, hijackings, shoot-downs)
- 4%: Other human error (air traffic controller error, improper loading of aircraft, improper maintenance, fuel contamination, language miscommunication etc.)
- 1%: Other cause

**2. Aircraft accident statistics** [2] provide a valuable source of information in order to set the proper priorities as well as to monitor progress made by the aviation industry. They are calculated for many kinds of aircraft. The accident statistics presented in this section apply to worldwide commercial jet airplanes that are heavier than 60,000 pounds maximum gross weight. However, airplanes manufactured in the former Soviet Union (CIS) are not included.

Aircraft accident statistics provide a valuable source of information in order to set the correct priorities as well as to monitor progress made by the aviation industry. They are calculated for many types of aircraft.

Globally, air traffic has increased since the early 20$^{th}$ Century and 2001. But after the World Trade Center attack on 9/11/01, many airlines were in financial difficulty, and worldwide air traffic started to decrease. Two years later growth recommenced.

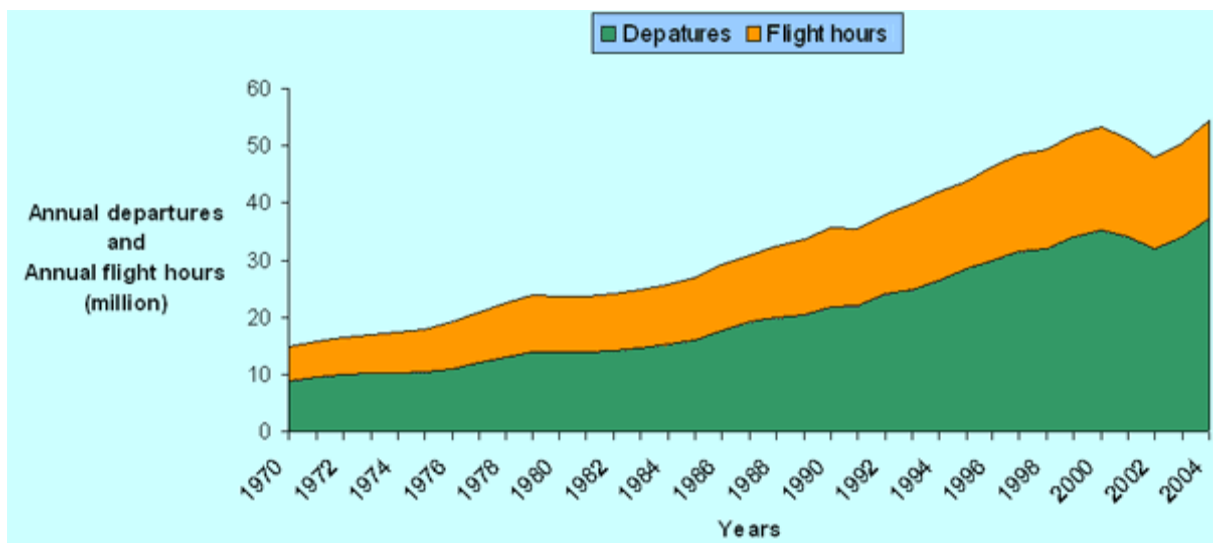

**Fig. 1**. Annual departures and flight hours (millions).

In 2004, airplanes flew 34.1 million flight hours. This number must be kept in mind for statistical analysis and failure rates interpretation: if an event has a probability to occur of 1 out of 1 million (which is rather low), it will occur several times a year.

The following graph shows the total number of certified commercial jet airplanes that are heavier than 60,000 pounds maximum gross weight (it does not include airplanes manufactured in the former Soviet Union).

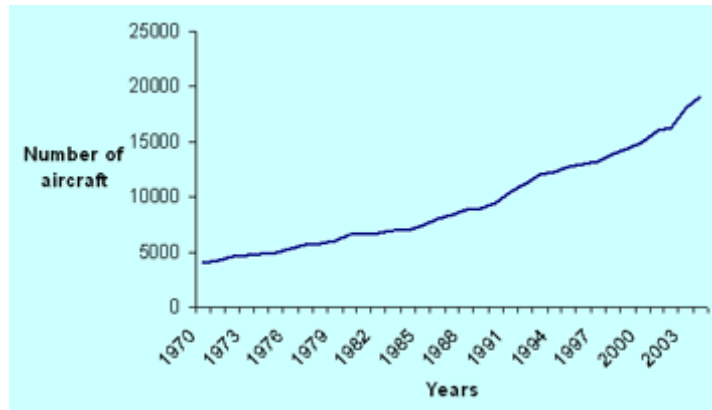

**Fig.2.** Number of aircraft in the World.

The number or aircraft ceaselessly increases to satisfy the growing transport demand by the public. Fortunately, even if the busiest airspaces are congested, the risk of collision is very low thanks to the new technologies that make possible an accurate position and altitude measurements (which definitely help both crews and Air Traffic Controllers).

For example, in order to accommodate all the traffic over some ocean routes, specific airspaces where aircraft are vertically separated by only 1000 ft were created. They are named RVSM airspaces, which stands for Reduced Vertical Separation Minima.

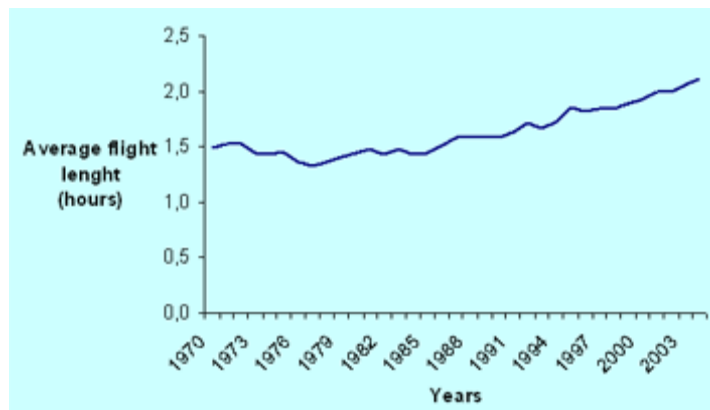

**Fig.3**. The average flight length (hours).

The graph above illustrates that the average flight time also increases with years. Some state-of-the-art aircraft can perform 22-hour non-stop flights.

**3. Aircraft accidents rate.** The number of accidents per flight decreases with time. But the number of fatalities per year is changing and does not decrease.

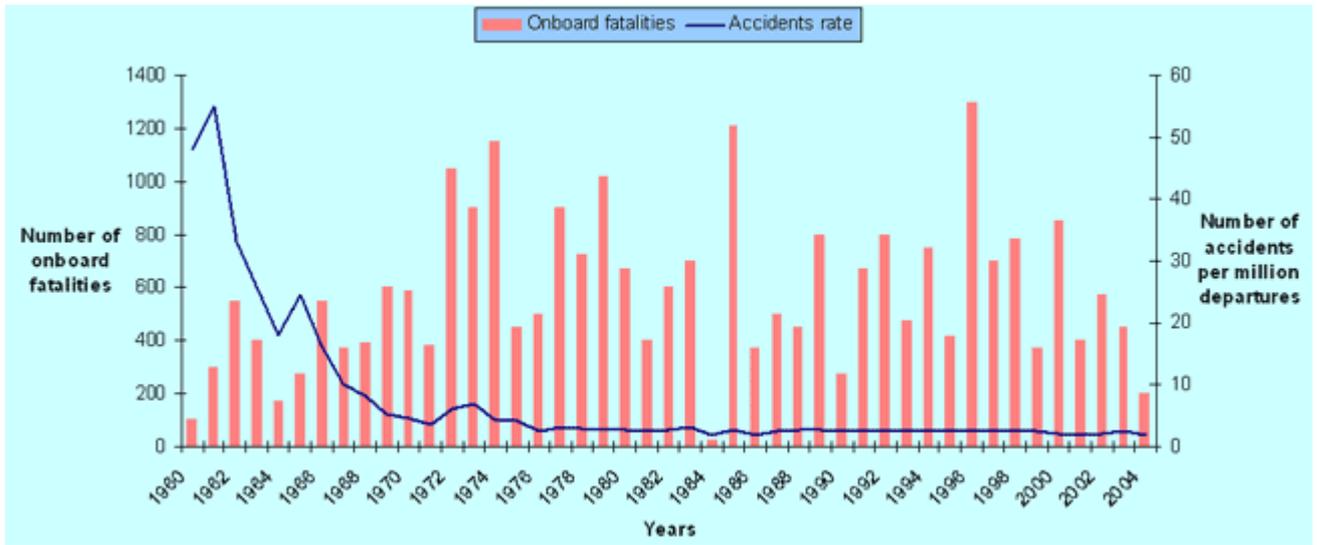

**Fig.4.** The Number of onboard fatalities and number of accidents per million departures.

Nowadays, aircraft accidents are less likely than several decades ago. Nevertheless, the growing number of flying aircraft and their increasing capacity cannot result in a reduction of onboard fatalities.

| Table 1 | 1959-2004 | 1995-2004 |
|---|---|---|
| Number of accidents | 1 402 | 376 |
| Number of onboard fatalities | 25 664 | 5 612 |

**4. Scheduled passenger operations are less likely to have an accident than other type of operation.** As newspapers always say, you are less likely to have an accident if you fly a regular flight than if you get on other types of flights (unscheduled passenger and charter, cargo, ferry, test, training, and demonstration). This is a truth of life you may also realize by examining the following graph. This graph also illustrates that scheduled passenger operations are 5 to 6 times more numerous.

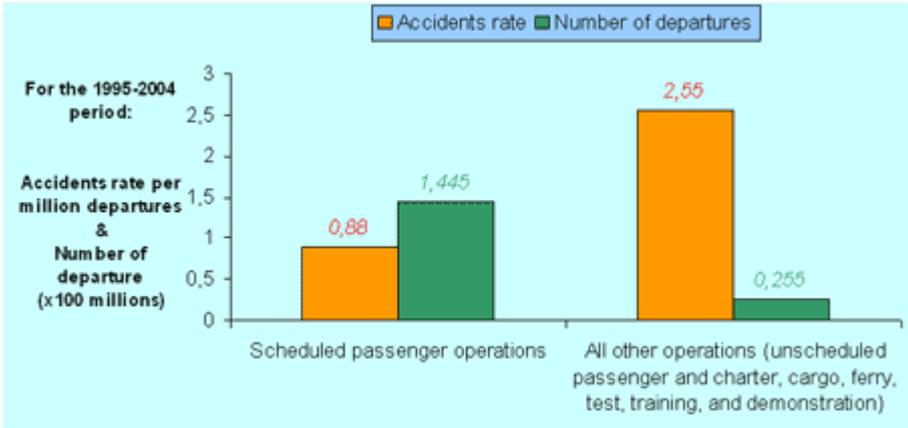

**Fig. 5.** Root causes of aircraft accidents.

It is quite exceptional for an accident to be related to one single cause. Almost every mishap is the consequence of a chain of events and accident reports usually make a distinction between the main cause and a number of contributing factors. The following graph shows the repartition of the main causes identified for aircraft crashes.

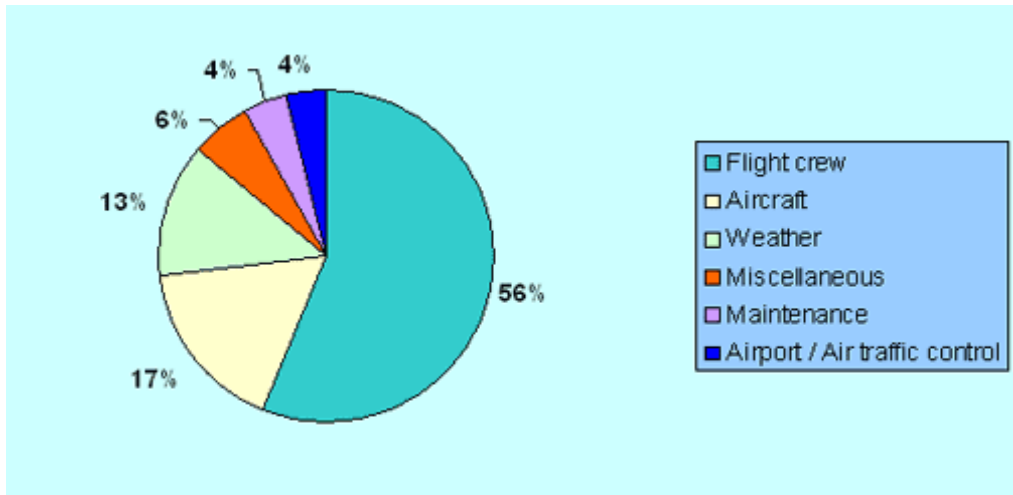

**Fig.6.** Percentage of different aircraft loss causes.

The main root cause is the infamous "human factors". In order to prevent this source of accidents, crews are requested to regularly re-train. Next comes the aircraft failures, but this cause is less likely with modern aircraft.

**5. Aviation terrorism**. Only airport security services and alert people can prevent aviation terrorism such as bombing and hijacking. Aircraft are very exposed to such attacks, because they are vulnerable, and because an aircraft crash is newsworthy. Nowadays, cargo bay containers are bomb-resistant, but not aircraft since such over-building would make them too heavy to fly!

Fortunately, and thanks to the numerous ideas that have been imagined to prevent such attacks, the number of terrorism acts has been reduced with years. But terrorism fatalities have no obvious trend. New aircraft carrying a lot of people (+600) may become a special terrorism target, which could lead to extremely deadly attacks in the next years.

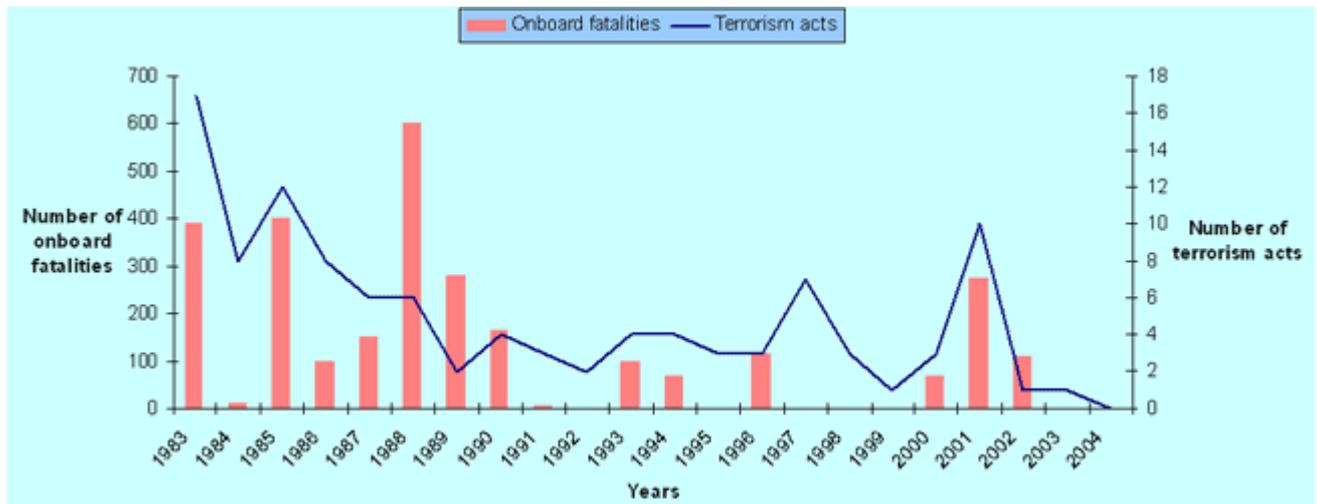

**Fig. 7.** Number of onboard fatalities and number of terrorism acts,

About 40 passengers out of 1 million are flight fatalities. That way some people avoid to flight by aircraft. We must decrease the flight death rate and give the hope to passengers: whatever occurrence the passengers have high chance to outlive the aircraft terrorism event.

# Innovations

In case of failure or full damage of main aircraft system, frame, wing, tail and control (for example, by terrorist rocket) for saving of passenger lives the pilot uses our system. The offered installation has the following peculiarities and devices:

1) **Small break parachute**. The parachute is a well-known device used by parachutists for air jumps and airborne re-supply aviation for airdropping of materials. But usually a parachute is not suitable for saving conventional air passenger and other freight aircraft. The conventional aircraft parachute must be very big, (that means - heavy). That gives a high landing speed. The aircraft has touchdown speed about 0.3 ÷ 1 m/s and the undercarriage brake distance is 0.5 ÷ 0.8 m. That way, the passenger feels only a small blow in the touchdown (landing) moment. The parachutist is trained man who is landing on his legs. His parachute has an air drag only about 10 N/m$^2$. However, parachutist has speed 5 ÷7 m/s and 5 ÷ 7 "g" overload in touchdown. The regular passenger is non-trained man (or women, child) who is sitting in aircraft chair. The parachutist lands on his legs and has brake distance 0.6 - 0.8 m. The regular passenger seats in aircraft chair that has the brake distance of 5-8 cm. That means the passenger overload is 50 ÷ 70 "g". That is a dead overload for any men, because the trained man can keep the maximum overload 16 "g" in a very short time period. That is full destroying of aircraft because the safety aircraft overload is 2 - 3 "g". We can save our regular passengers and aircraft if the touchdown speed will be close to zero! There were a lot of projects for saving small aircraft by parachutes [3], but most of them were useless because even big (and heavy) parachute cannot basically make small touchdown speed and save passenger and aircraft.

As it is noted, we use a small brake which decreases the speed only for 25 ÷ 40 m/s. The parachute drag depends from speed as $V^2$. That means our parachute will be in 25 ÷ 100 times less (and easier weight) than conventional parachute used in other projects. (How to decrease speed from 20 - 40 m/s to 0 m/s, you see in next point). We can use the inflatable wing parachute having ratio lift/drag equals 6 - 7 and glide in long distance.

The other difficult is aircraft overload in a moment of canopy opening. If aircraft speed is high, the overload is very big. We use a special parachute (or double) which canopy opens slowly with permanent overload (2 ÷ 3g). This saves passengers and the aircraft.

The next problem is control. The pilot must have a choice in a landing place out of building, high voltage electric power-lines, trees, etc. We use the control 'square' ram-air parachute (airfoil, parafoil) that gives a limited control in choice of landing place.

At present the time there are available strong, light artificial fibers (materials) which can be used for offered parachute [4] p. 33.

**Table 2.** Tensile strength and density of whiskers and fibers

| Material Whiskers | Tensile strength kgf/mm$^2$ | Density g/cm3 | Fibers | Tensile strength kgf/mm$^2$ | Density g/cm$^3$ |
|---|---|---|---|---|---|
| AlB$_{12}$ | 2650 | 2.6 | QC-8805 | 620 | 1.95 |
| B | 2500 | 2.3 | TM9 | 600 | 1.79 |
| B$_4$C | 2800 | 2.5 | Thorael | 565 | 1.81 |
| TiB$_2$ | 3370 | 4.5 | Alien 1 | 580 | 1.56 |
| SiC | 2100-4140 | 3.22 | Alien 2 | 300 | 0.97 |
| Al oxide | 2800-4200 | 3.96 | Kevlar | 362 | 1.44 |

See Reference [4] p. 33.

The fibrous material PRD-49 had the following properties in 1972 [5]:
Tensile ult. 400 psi (312 kgf/mm$^2$); specific gravity 1500 kg/m$^3$; filament diameter 0.0003 in.; minimum available yarn size (denier) 200.

Summary our parachute innovation's features:
a) We use small, light parachutes which decreases aircraft speed only up $V = 20 ÷ 40$ m/s.
b) Our parachute opens with the constant overload which is safe for passengers and aircraft.
c) Our parachute is ram-air and has limited control.

d) It is made from currently availabe commercial artificial fibers that are strong.
e) Our parachute is outer device and can be joined to (or disconnected from) any existing aircraft to main frame or special round type (Fig. 8).

**2. Rocket booster**. For decreasing (braking) the touchdown speed to zero solid fuel rockets are used. The change of speed (20 - 40 m/s) is small and rocket booster is also small. Its weight is 1 - 2% of aircraft weigth. The small solid rocket engine is simplest, reability, cheap. Its combustion camera (body) may be made from the current composed material and has a small weight (10 - 15% from fuel weight). The rocket booster is packed with parachute and take out together with them. It is located under open parachute and works authomatically only distance to ground is small (see Fig. 8).

**3. Aircrcraft inflateble pillows.** For soft landing and saving passenger in landing to sea, our system may to have inflatable pillows located below of fugelage in special low drag containers. They are filling an air before the aircraft touchdown and save aircraft in any case when one is landing on fuselage (not gears) or in water. Their weight is small (see project).

The offered method has very important advantage: in emergency situation the aircraft can make vertical landing in any place without demage.

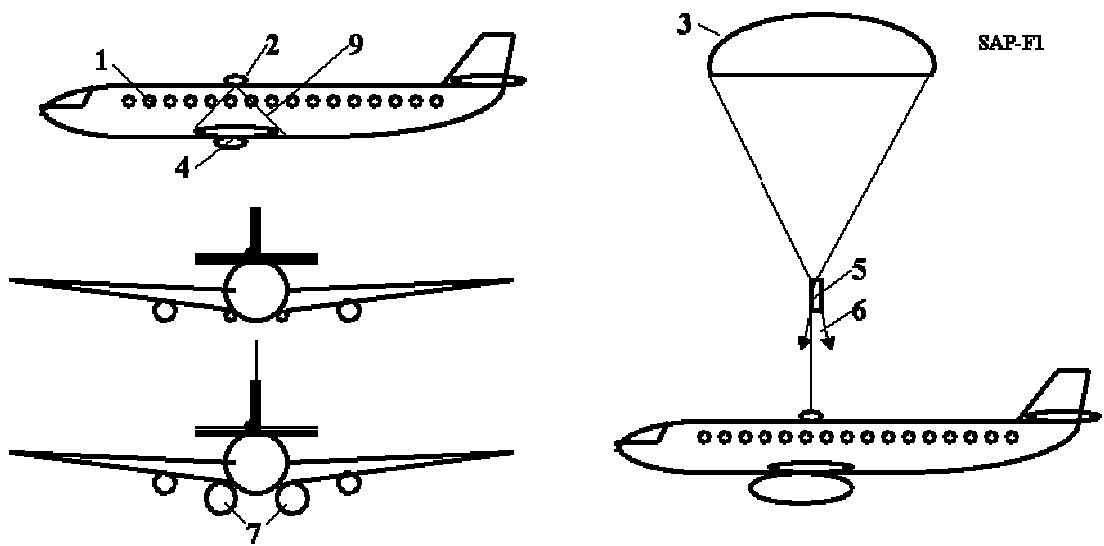

**Fig. 8**.
Method for saving passenger lives in any aircraft demagas. Notations: 1 - aircraft, 2 - parachute booster container, 3 - parachute, 4 - inflatable pillow-float container, 5 - rocket booster, 6 - booster jet, 7 - pillow-float in swollen form, 9 - support type.

## Computations and Estimations

1. **Parachute drag**. Relative parachute drag compute by equation

$$\overline{D} = \frac{D}{S} = C_D \frac{\rho V^2}{2}, \qquad (1)$$

where $D$ is air drag, N/m$^2$; $S$ is parachute area, m$^2$; $C_D = 1 \div 1.6$ is drag coefficient; $\rho = 1.225$ is air density, kg/m$^3$; $V$ is parachute speed, m/s. Computations are presented in Fig.9.

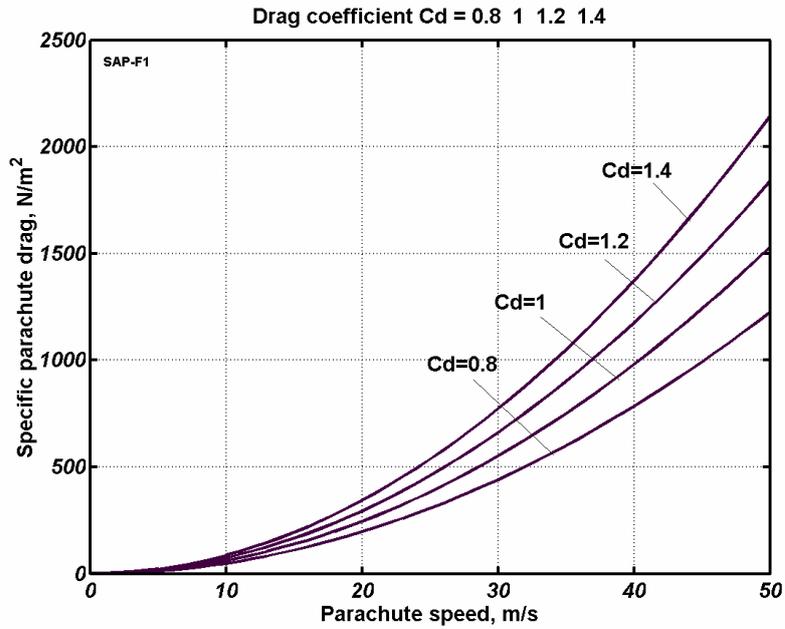

**Fig.9.** Specific parachute drag (N/m$^2$) via parachute speeds for different parachute drag coefficient.

**2. Parachute parameters**. The equation below allow to compute the mass of round parachute

$$S = \pi R^2, \quad S_c = 2S, \quad \delta = \frac{\overline{D}R}{2\sigma}, \quad M = 2\pi R^3 \overline{D} \frac{\gamma}{\sigma},$$
$$for \ C_D = 1 \ \ M = 2\pi R^3 \frac{\rho V^2}{2} \frac{\gamma}{\sigma} \tag{2}$$

where $R$ is parachute radius, m; $\delta$ is parachute thickness; $\sigma \approx 150 \times 10^7$ is safety stress of parachute material, N/m$^2$; $M$ is parachute mass, kg; $\gamma \approx 1500$ kg/m$^3$ is density of parachute material, kg/m$^3$. Coefficient 2 means that weight of parachute cord apriximately equals the weight of parachute canope. The computations are shown in Fig.10.

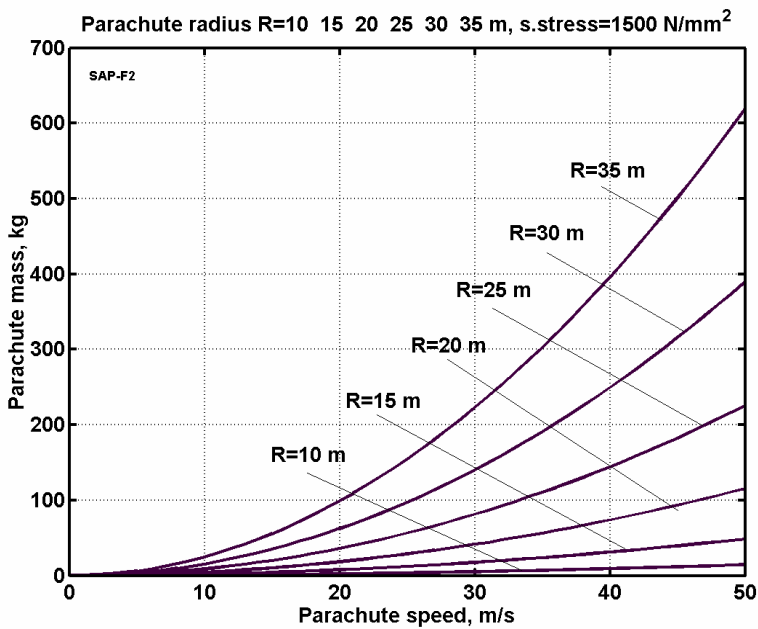

**Fig. 10**. Parachute mass via parachute speed for different parachute radius. $C_D = 1$.

**3. Mass of rocket booster.** The fuel mass of solid rocker booster may be estimated by equation

$$\overline{M}_b = \frac{V}{v}, \tag{3}$$

where $v \approx 1800 \div 2600$ m/s is specific impulse (jet speed) of rocket booster. Results of computation is presented in Fig.11. The body mass of booster rocket is about $10 \div 15\%$ for current composite material.

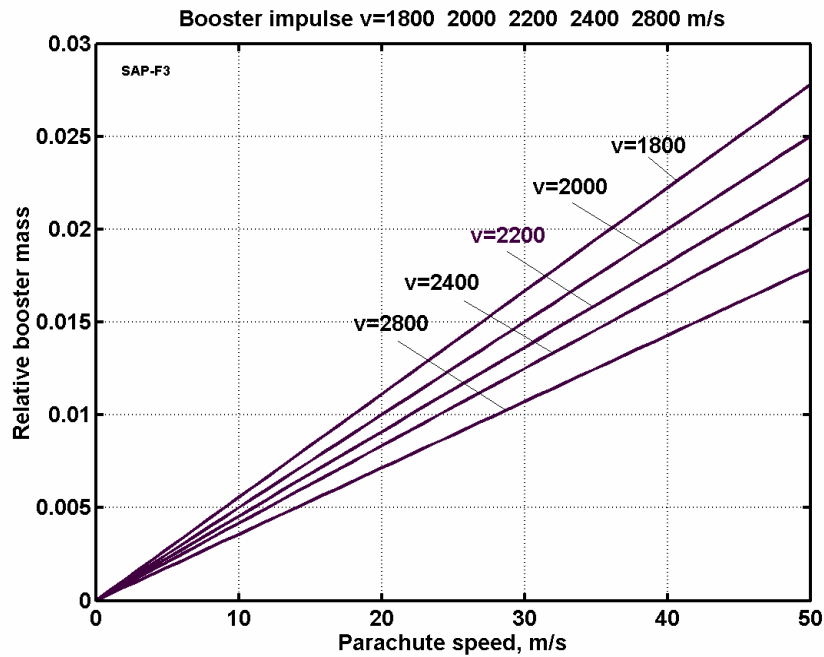

**Fig.11**. Relative booster mass via parachute speed for different booster fuel.

## Project

As example, we calculate the parachute paramethes for aircraft having weight $W = 100$ tons ($2\times10^6$ N/m$^2$). Take as initial data: the parachute speed $V = 30$ m/s, drag coefficient $C_D = 1.2$, specific impulse of booster $v = 2200$ m/s. The requested area of parachute is

$$S = \frac{2W}{C_D \rho V^2} = \frac{2\times 10^6}{1.2 \cdot 1.225 \cdot 30^2} = 1500 \text{ m}^2, \quad R = \sqrt{\frac{S}{\pi}} = 21.8 \text{ m}^2.$$

Canope tickness and total mass of parachute for $\sigma = 150\times 10^7$ N/m$^2$, $\gamma = 1500$ kg/m$^2$ are

$$q = \frac{\rho V^2}{2} = 551, \quad \delta = C_D q \frac{R}{2\sigma} = \frac{551 \cdot 21.8}{2 \cdot 1.5 \cdot 10^9} = 4\cdot 10^{-6} \text{ m},$$

$$M = 4S\delta\gamma = 4 \cdot 1500 \cdot 4 \cdot 10^{-6} \cdot 1500 = 36 \text{ kg}$$

Mass of rocket booster for $v = 2200$ m/s is

$$\overline{M}_b = 1.15\frac{V}{v} = 1.15\frac{30}{2200} = 0.0157.$$

Where weight of booster body is 15%.
Mass of air pillows together with air balloon is about 14 kg, support types 20 kg.
The total mass of installation is 1640 kg. We take 1700 kg. That is 1.7% from total mass of aircraft. That is small prize for saving passenger lives in any catastrofic situation.

## Conclusion

Offered is an entirely new method of saving passenger lives in any catastrofic situation including full failure control, damage and loss of all aircraft wings, tail, breakdown of all engines, etc. It shown that previous works proposed using only jet parachute are useless because they did not consider the overload of the parachute jerk (in release parachute) and the blow (touchdown) of aircraft on the Earth's surface. It is offered the series of innovations which overcome these difficulties and allow a soft landing in any suitable place, allowing preservation of the aircraft. This method may be applied to all existing airplanes and increases their weight only in 1.5 ÷ 2.5%. This method may be also used for vertical landing aircraft, for example, when airport runway is damaged or overloaded.

See also the works [6]-[7].

## Acknowledgement

The author wishes to acknowledge R.B. Cathcart for correcting the author's English and useful advice.